\documentclass[aps,pre,epsf,superscriptaddress,amsmath,amssymb,amsfonts,twocolumn,showpacs]{revtex4-1}

\usepackage{graphicx}
\usepackage{epsfig}
\usepackage{dcolumn}
\usepackage{bm}
\usepackage{braket}
\usepackage{amsmath}
\usepackage{mathtools}
\usepackage{color}
\newcommand{\abs}[1]{\left| #1 \right|}

\begin{document}
\title{Repulsive Fermi Polarons and Their Induced Interactions\\
in Binary Mixtures of Ultracold Atoms}

\author{S.I. Mistakidis}
\affiliation{Center for Optical Quantum Technologies, Department of Physics, University of Hamburg, 
Luruper Chaussee 149, 22761 Hamburg Germany}

\author{G.C. Katsimiga}
\affiliation{Center for Optical Quantum Technologies, Department of Physics, University of Hamburg, 
Luruper Chaussee 149, 22761 Hamburg Germany}  

\author{G.M. Koutentakis}
\affiliation{Center for Optical Quantum Technologies, Department of Physics, University of Hamburg, 
Luruper Chaussee 149, 22761 Hamburg Germany}\affiliation{The Hamburg Centre for Ultrafast Imaging,
Universit\"{a}t Hamburg, Luruper Chaussee 149, 22761 Hamburg,
Germany}

\author{P. Schmelcher}
\affiliation{Center for Optical Quantum Technologies, Department of Physics, University of Hamburg, 
Luruper Chaussee 149, 22761 Hamburg Germany} \affiliation{The Hamburg Centre for Ultrafast Imaging,
Universit\"{a}t Hamburg, Luruper Chaussee 149, 22761 Hamburg,
Germany}

\date{\today}

\begin{abstract}
We explore repulsive Fermi polarons in one-dimensional harmonically trapped 
few-body mixtures of ultracold atoms using as a case example a $^6$Li-$^{40}$K mixture. 
A characterization of these quasiparticle-like states,
whose appearance is signalled in the impurity's 
radiofrequency spectrum, is achieved by extracting 
their lifetime and residua.
Increasing the number of $^{40}$K impurities
leads to the occurrence of both single and multiple polarons that are entangled with their environment. 
An interaction-dependent broadening of the spectral lines
is observed suggesting the presence of induced interactions.
We propose the relative distance between the impurities as an adequate measure 
to detect induced interactions independently of the specifics of the atomic mixture, 
a result that we showcase by considering also a $^6$Li-$^{173}$Yb system. 
This distance is further shown to be indicative of the generation of entanglement independently of the size of the 
bath ($^6$Li) and the atomic species of the impurity. 
The generation of entanglement and the importance of induced interactions are revealed 
with an emphasis on the regime of intermediate interaction strengths. 
\end{abstract}

\maketitle

\section{Introduction} 

The properties and interactions of impurities immersed in a complex many-body (MB) environment
represents a famous example of Landau's quasiparticle theory~\cite{Landau}. 
The concept of a polaron, where an impurity immersed in a bath couples to the excitations of the latter forming 
an effective free particle, plays a central role in our understanding of quantum matter. 
Applications range from semiconductors~\cite{Gershenson}, high $T_c$ superconductors~\cite{Dagotto},
and liquid Helium mixtures~\cite{Bardeen,Baym} to polymers and proteins~\cite{Deibel,Davydov}.
Population imbalanced ultracold Fermi gases \cite{Pecak} with their tunable interactions,
offer an ideal platform for studying the impurity problem 
as well as the effective interactions between Fermi polarons.

Most of the experimental and theoretical studies on this topic have 
initially been focusing on attractive Fermi polarons~\cite{Schirotzek,Navon,Punk,Chevy,Zhang,Tajima}. 
Only very recently quasiparticle formation in fermionic systems associated with strong repulsive interactions 
have been experimentally realized first in the context of narrow~\cite{Kohstall} 
and subsequently for universal broad Feshbach resonances~\cite{Scazza,Koschorreck}. 
They have triggered a new era of theoretical investigations regarding the properties of repulsive Fermi 
polarons~\cite{Cui,Pilati,Massignan1,Schmidt1,Schmidt2,Ngampruetikorn,Massignan2,Schmidt3}. 
These metastable states--that can decay into molecules in two- and 
three-dimensions (3D)--are of fundamental importance 
since their existence and longevity offers the possibility of stabilizing repulsive Fermi gases. 
As a result exotic quantum phases and itinerant 
ferromagnetism~\cite{Duine,Chang,Pekker,Sanner,He,Valtolina,Li,Georgetos} could be explored. 
While for finite impurity mass Fermi polarons constitute well-defined quasiparticles in these higher dimensional
systems~\cite{Zoellner,Meera,Bour}, the quasiparticle picture is shown to be ill-defined in the thermodynamic limit of 
one-dimensional (1D) settings~\cite{Rosch,Castella,Massignan3}.
However, important aspects of the physics in this limit have been identified in few-body 
experiments evading such difficulties~\cite{Brouzos,Gharashi}. 
Besides the fundamental question of the existence of coherent quasiparticles in such lower dimensional 
settings~\cite{Gharashi,Yang,Mathey,Leskinen,Catani,Casteels,Doggen,Mao,Parisi,Pastukhov,artem2,Mistakidis4,Mistakidis5}, 
far less insight is nowadays experimentally available regarding the notion of induced 
interactions between polarons~\cite{Klein,Recati,Mora,comment,Yu,Kinnunen,Hu,Dehkharghani,Grusdt,Guardian,Naidon,Guardian1}. 
In this direction, 1D systems represent the cornucopia for studying effective interactions between  
quasiparticle-like states, since their role is expected to be enhanced in such settings~\cite{Massignan3}. 

In this work, we simulate the experimental process of reverse radiofrequency (rf) 
spectroscopy~\cite{Kohstall,Scazza,Gupta,Regal} using as a case example a mixture consisting of $^{40}$K Fermi 
impurities coupled to a few-body $^{6}$Li Fermi sea 
and demonstrate the accumulation of polaronic properties. 
We predict and characterize the excitation spectrum of these states and derive their lifetimes and residua. 
Most importantly here, we identify all the dominant microscopic mechanisms that lead to polaron formation. 
By increasing the number of $^{40}$K impurities immersed in a $^6$Li bath, 
we verify the existence of single as well as multi-polaron states both for weak 
and strong interspecies repulsion. 
In line with recent studies~\cite{Scazza,Cetina} the presence of induced interactions between the polarons is indicated 
by a positive resonance shift further accompanied by a spectral broadening. 
However, the non-sizeable nature of this shift, being of the order of $2\%$, 
suggests that in order to infer about the presence of induced interactions an alternative 
measure is needed. Inspecting the relative distance between the resulting
quasiparticles, a quantity that can be probed experimentally via {\it in-situ} spin-resolved 
single-shot measurements~\cite{Jochim2}, we observe its decrease which concordantly 
dictates the presence of induced interactions~\cite{Jie}. 
The latter are found to be attractive despite the repulsive nature of the fundamental interactions in the system. 
This fact persists upon enlarging the fermionic sea~\cite{comment} and considering different atomic species. 
We find that the decrease of the relative distance is inherently connected to the generation of entanglement. 
The von-Neumann entropy~\cite{Horodeki} reveals equally the entanglement and is sensitive to the number of
impurities. 

This work is structured as follows. 
Section \ref{theory} presents our setup and many-body treatment. 
In Section \ref{results} we discuss the excitation spectrum of the fermionic mixture and identify 
polaronic states. 
We also extract their residues and lifetimes. 
In Section \ref{entanglement} we quantify the degree of entanglement between the impurity atom(s) and the bath. 
The induced interactions between the two impurity atoms are analyzed and related to the generation of entanglement. 
We summarize our findings and provide an outlook in Sec. \ref{conclusions}. 
Appendix \ref{rf_implement} contains a discussion of our numerical implementation regarding the process of rf spectroscopy. 
The applicability of the employed model in the context of effective range corrections is shown in Appendix \ref{sec:1d}. 
Appendix C showcases the behavior of the energy of the polaron versus the particle number of the bath. 
Finally, in Appendix \ref{sec:numerics1} we provide further details of our numerical findings presented in the main text.

\begin{figure*}[ht]
\includegraphics[width=1.0\textwidth]{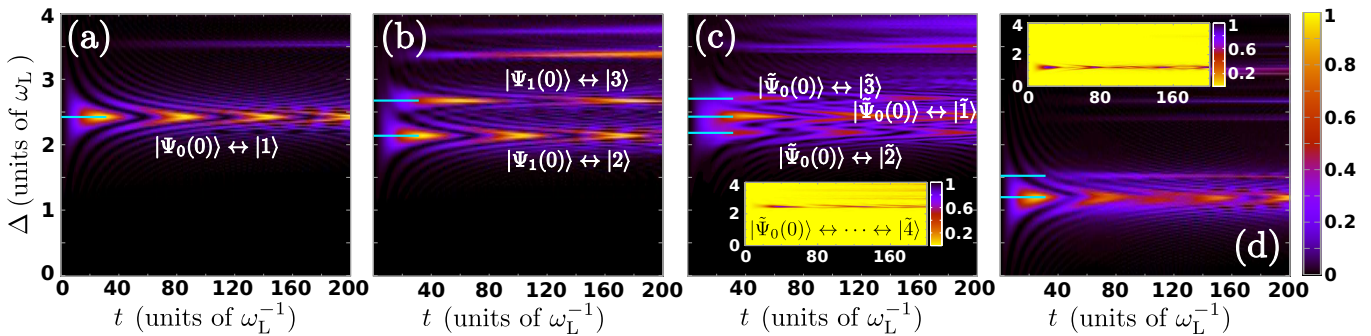}
\caption{(Color online) Spectroscopic signal, $f(\Delta, t)$, of a single impurity initialized (a) in the ground state,
and (b) in the first excited state of the harmonic oscillator for 
strong interspecies repulsive interactions, $g=5$, verifying the existence of well-defined quasiparticle peaks that appear in the rf spectrum. 
(c) Same as (a) but for two impurities.
(d) Same as (c) but for weak coupling ($g=1.5$). 
The insets in (c), (d) present the probability of finding two spin flipped impurities.
Markers in light blue indicate the center of each resonance $\Delta_{+_i}$. } 
\label{Fig:1}
\end{figure*}

\section{Theoretical Framework} \label{theory}

\subsection{Model System} \label{setup}

Our system consists of $N_{L}=5$ spinless $^{6}$Li fermions 
each with mass $m_{L}$, which serve as a bath for the spin-$1/2$ $N_{K}=1,2$ $^{40}$K impurities of mass $m_{K}$. 
Each species is trapped in a 1D harmonic potential with frequency $\omega_{K}=0.6~\omega_{L}$
in line with previous $^{6}$Li-$^{40}$K experiments~\cite{Tiecke,Naik,Cetina0,Cetina}.   
The MB Hamiltonian of the system reads 
\begin{eqnarray}
\hat{H} = \hat{H}^{0}_{L}+\sum_a \hat{H}^{0}_a+\hat{H}_I+\hat{H}_S, 
\label{Htot}
\end{eqnarray}
where $\hat{H}^{0}_{L}=\int dx~\hat{\Psi}^{\dagger}_{L} (x) \left( -\frac{\hbar^2}{2 m_{L}} \frac{d^2}{dx^2}  
+\frac{1}{2} m_{L} \omega_{L}^2 x^2 \right) \hat{\Psi}_{L}(x)$, 
is the Hamiltonian describing the trapped motion of the majority $^6$Li atoms with trap frequency $\omega_{L}$. 
The corresponding non-interacting Hamiltonian of the minority $^{40}$K atoms is 
$\hat{H}^{0}_a=\int dx~\hat{\Psi}^{\dagger}_a \left( -\frac{\hbar^2}{2 m_K} \frac{d^2}{dx^2}  
+\frac{1}{2} m_K \omega_K^2 x^2 \right) \hat{\Psi}_a(x)$, where $a=\left\{ \uparrow, \downarrow \right \}$ denotes the spin component.   
In both of the above-mentioned cases $\hat{\Psi}_i (x)$ is the fermionic field-operator for either the majority ($i=L$) 
or the impurity ($i=K$) atoms.  
The contact interspecies interaction term of effective strength $g>0$ between a spin-$\uparrow$ 
impurity particle and the bath is given by
$\hat{H}_I=g \int dx~\hat{\Psi}^{\dagger}_{L}(x) \hat{\Psi}^{\dagger}_{\uparrow}(x) \hat{\Psi}_{\uparrow} (x)\hat{\Psi}_{L}
(x)$~\cite{g1d}. 
The non-resonant interaction of the spin-$\downarrow$ state with the $^{6}$Li bath can be neglected 
when compared to $\hat{H}_I$. 
Moreover, the effective interaction strength $g$ \cite{olsani} can be experimentally tuned either by means of 
Feshbach resonances \cite{Chin} or confinement induced resonances \cite{olsani}. 
It is important to stress at this point that a bound state of a Feshbach molecule occurs for all 
scattering lengths in one-dimension. 
However it can be demonstrated \cite{Heidelberg3,Frank} that its effect is negligible for repulsive interactions sufficiently 
away from the infinite interaction limit such as the ones considered herein (see also Appendix A). 
We also note that in the considered few-body case it can be shown that the effective range corrections to 
the interaction term $\hat{H}_I$ stemming from the presence of narrow Feshbach resonances 
are negligible, see Appendix \ref{sec:1d}. 
Finally, $\hat{H}_S=\frac{\hbar \Omega^0_R}{2} \hat{S}_x - \frac{\hbar \Delta}{2} \hat{S}_z$,
where $\Omega^0_R$ denotes the Rabi frequency, and $\Delta$ the detuning 
of the rf field in the absence of the $^6$Li bath.
Here, $\hat{\textbf{\textit{S}}}=\int dx \sum_{ab} \hat{\Psi}_a (x) \text{\boldmath$\sigma$}_{ab} \hat{\Psi}_b (x)$
is the total spin operator while \text{\boldmath$\sigma$} denotes the Pauli vector. 
We assume, $\Omega^0_R\ll \omega_{L}$ such that $\Omega^0_R\ll \Delta_{+_i}$ 
($\Delta_{+_i}$ denotes the location of the resonance to the $i$-th state identified in the rf spectra) 
thus allowing for a spectroscopic study of the polaronic structures, see also Appendix \ref{rf_implement}.

\subsection{The Many-Body Approach} \label{sec:many_body_ansatz}

To theoretically address the impurity problem, we use a variational method, namely
the Multi Layer Multi-Configuration Time-Dependent Hartree method for atomic mixtures (ML-MCTDHX),
that takes into account all particle correlations~\cite{Alon,moulosx}.
Such a non-perturbative inclusion of correlations allows us to calculate the impurity spectrum
and thus identify the emergent polaron states. 

The MB wavefunction, $|\Psi(t)\rangle_{MB}$ is constructed as a linear combination
of a set of $M$ time-dependent wavefunctions for each of the species being referred to as species
wavefunctions, $|\Psi^{\sigma}_i(t)\rangle$. Here $\sigma \in \{ {\rm L}, {\rm
K} \}$, $i=1,\dots,M$ and 
\begin{equation} 
    |\Psi(t)\rangle_{MB}=\sum_{i,j=1}^M A_{ij}(t) |\Psi^{\rm L}_i(t)\rangle|\Psi^{\rm K}_j(t)\rangle,
    \label{eq:tot_wfn}
\end{equation}
where $A_{ij}(t)$ denote the time-dependent expansion coefficients. 
Equation (\ref{eq:tot_wfn}) is equivalent to a truncated Schmidt decomposition of rank $M$ \cite{Horodeki,darkbright,phassep}.  
Indeed, the spectral decomposition of the expansion coefficients $A_{ij}$ reads 
$A_{ij}(t)=\sum_{k=1}^M U_{ik}^{-1}(t)\sqrt{\lambda_k(t)}U_{kj}(t)$ where $\sqrt{\lambda_k(t)}$ refer to the Schmidt weights. 
As a consequence the MB wavefunction can be written as a truncared Schmidt decomposition i.e. 
$|\Psi(t)\rangle_{MB}=\sum_{k=1}^M \sqrt{\lambda_k(t)} |\tilde{\Psi}^{\rm L}_k(t)\rangle|\tilde{\Psi}^{\rm K}_k(t)\rangle$.  

Subsequently each of the species wavefunctions is expanded on the time-dependent
number-state basis, $|\vec{n} (t) \rangle^{\sigma}$, with time-dependent weights 
$B^{\sigma}_{i;\vec{n}}(t)$ 
\begin{equation}
    | \Psi_i^{\sigma} (t) \rangle =\sum_{\vec{n}} B^{\sigma}_{i;\vec{n}}(t) | \vec{n} (t) \rangle^{\sigma}.
    \label{eq:mb_wfn}
\end{equation}
Each time-dependent number state corresponds to a Slater determinant of the $m^{\sigma}$ time-dependent
variationally-optimized single-particle functions (SPFs) $\left|
\phi_l^{\sigma} (t) \right\rangle$, $l=1,2,\dots,m^{\sigma}$ with occupation
numbers $\vec{n}=(n_1,\dots,n_{m^{\sigma}})$. 
Each of the SPFs is subsequently expanded in a primitive basis. 
For the ${}^{6}$Li atoms the primitive basis $\lbrace \left| k \right\rangle \rbrace$ consists of a discrete variable
representation (DVR) of dimension $\mathcal{M}$. For ${}^{40}$K
the primitive basis $\lbrace \left| k,s \right\rangle \rbrace$, refers to the
tensor product of a the aforementioned DVR
basis for the spatial degrees of freedom and the two-dimensional spin basis $\{
\ket{\uparrow}, \ket{\downarrow}\}$,
\begin{equation}
    | \phi^{\rm K}_j (t) \rangle= \sum_{k=1}^{\mathcal{M}} \sum_{\alpha=\{ \uparrow, \downarrow \}} C^{{\rm
    K}}_{jk \alpha}
    (t) \ket{k} \ket{\alpha}.
    \label{eq:spfs}
\end{equation}
$C^{{\rm K}}_{j k \alpha}(t)$ refer to the corresponding time-dependent expansion
coefficients.  Note here that each time-dependent SPF for the ${}^{40}$K is a general
spinor wavefunction of the form $| \phi_j^{\rm K} (t) \rangle= \int {\rm d}x~ [
    \chi_j^\uparrow(x) \hat{\Psi}^\dagger_\uparrow(x) + \chi_j^\downarrow(x)
\hat{\Psi}^\dagger_\downarrow(x) ] | 0\rangle$ (see also \cite{Georgetos}). 
The time-evolution of the $N$-body wavefunction under the effect of the Hamiltonian $\hat{H}$ reduces to
the determination of the $A$-vector coefficients and the expansion coefficients of each
of the species wavefunctions and SPFs. 
Those, in turn, follow the variationally obtained ML-MCTDHX equations of motion \cite{moulosx}. 
It is important to mention here that in order obtain the eigenstates involving one and two polarons of the interacting MB system 
we use the method of improved relaxation~\cite{moulosx} within ML-MCTDHX.
We remark that the system in its stationary state reduces to a binary mixture of bath and spin-$\uparrow$ atoms.
In this way the general ansatz of Eq.~(\ref{eq:tot_wfn}) becomes that of Eq.~(\ref{schmidt}) [see also the discussion in Sec. \ref{entanglement}].
For a detailed discussion on this ansatz we refer the reader to~\cite{moulosx,darkbright,phassep}.  

In the limiting case of $M=1$ and $m^{\sigma}=N^{\sigma}$ the method reduces to the two-species coupled time-dependent
Hartree-Fock method, while for the case of $M=\min\left[\binom{m^{\rm
L}}{N^{\rm L}},\binom{m^{\rm K}}{N^{\rm K}} \right]$, $m^{\rm
L}=\mathcal{M}$ and $m^{\rm K}=2 \mathcal{M}$, it is equivalent to a full
configuration interaction approach (commonly referred to as ``exact
diagonalization'' in the literature) within the employed primitive basis. 
Another important reduction of the method is the so-called species mean-field (SMF) approximation \cite{moulosx,darkbright}. 
In this context the entanglement between the species is ignored while the correlations within each of the species 
are taken into account. 
More specifically, the system's wavefunction is described by only one species wavefunction, 
i.e. $\ket{\Psi_i^{L}(t)}=\ket{\Psi_{i}^{K}(t)}=0$ for $i\neq 1$. 
Subsequently each species wavefunction is expressed in terms of the time-dependent number state basis 
of Eq. (\ref{eq:mb_wfn}) consisting of different time-dependent variationally optimized SPFs. 
As a result the total wavefunction of the system takes the tensor product form 
\begin{equation} 
    |\Psi(t)\rangle_{SMF}= |\Psi^{\rm L}_1(t)\rangle \otimes |\Psi^{\rm K}_1(t)\rangle. 
    \label{eq:SMF}
\end{equation}

\begin{figure*}[ht]
\includegraphics[width=1.0\textwidth]{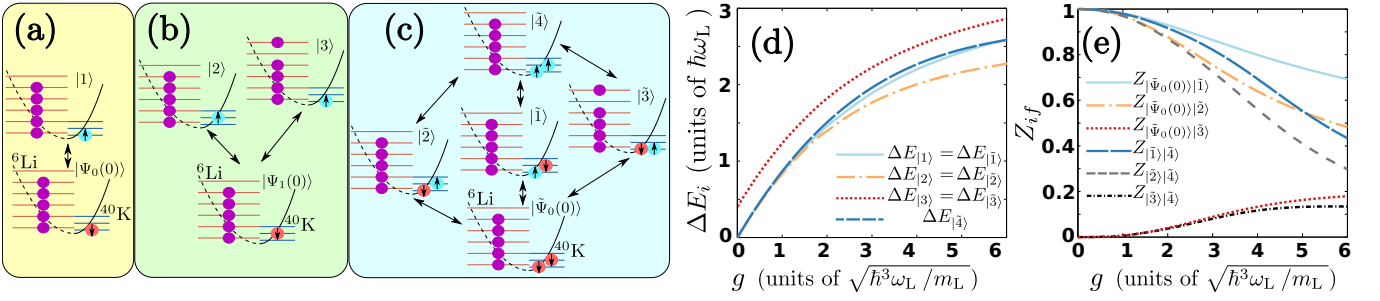}
\caption{(Color online) 
Schematic representation of the identified spectral transitions for (a), (b) $N_K=1$ and (c) $N_K=2$
$^{40}$K impurities immersed in the $N_{L}=5$ $^6$Li Fermi sea.
(d) Polaron energy branches, $\Delta E_i(g)$ for the different $i=\ket{1},\dots,\ket{4}$
identified transitions (see text).
(e) Residua, $Z_{fi}$, of the repulsive polarons calculated for varying $g$ and for each of the aforementioned transitions.} 
\label{Fig:2}
\end{figure*}

\section{Reverse Radiofrequency Spectroscopy} \label{results} 

\subsection{Excitation Spectrum}

In order to probe the excitation spectrum of the $^{40}$K impurities 
we simulate reverse rf spectroscopy~\cite{Kohstall,Scazza,Gupta,Regal}. 
This process follows the protocol explicated below. 
The initial state of the system consists of the $^6$Li atoms 
in their $N_{L}$-body non-interacting ground state 
$\ket{\Psi_{L}(0)}=\prod^{N_{L}-1}_{i=0} \int dx~\phi^{L}_i(x) \hat{\Psi}^{\dagger}_{L}(x) \ket{0}$. 
$\phi^{L}_i(x)$ refers to the $i$-th energetically excited eigenstate of $\hat{H}^0_{L}$. 
For $N_K=1$ the $^{40}$K impurity is prepared in the non-interacting spin-$\downarrow$ state,
and it is either in its ground state or in its first excited state (see also the discussion below).
Namely $\ket{\Psi_{j}(0)}=\int dx~\phi^{K}_j(x) \hat{\Psi}^{\dagger}_{\downarrow}(x) \ket{\Psi_{L}(0)}$,
where $j\in \{0, 1\}$ while $\phi^{K}_j(x)$ refer to the eigenstates of $\hat{H}^0_{\downarrow}$.
We then drive the impurity atom to the resonantly interacting spin-$\uparrow$ state, by applying a
rectangular rf pulse (see also Appendix \ref{rf_implement}) with bare Rabi frequency $\Omega^0_R=4 \pi \times 10^{-2}$ 
(harmonic oscillator units $\hbar=m_{L}=\omega_{L}=1$ are adopted here).
Our simulated spectroscopic signal presented in Figs.~\ref{Fig:1}(a), \ref{Fig:1}(b) 
is the fraction of impurity atoms transferred after a pulse $f(\Delta,t)=\frac{\langle N_{\uparrow} \rangle}{N_K}$, 
with $\langle N_{\uparrow} \rangle$ being the number of spin flipped impurities, 
measured for varying rf detuning $\Delta=\nu_{rf}-\nu_0$ and pulse time $t$.
$\nu_0$ denotes the frequency of the non-interacting transition between the 
spin-$\downarrow$ and spin-$\uparrow$ states and $\nu_{rf}$ is the applied 
frequency (see also Appendix \ref{rf_implement}).

Starting from $\ket{\Psi_{0}(0)}$ and for fixed strong interspecies repulsions ($g=5$) 
we observe a resonance for $\Delta_{+_{\ket{1}}}=2.430 \pm 0.002$ [see Fig.~\ref{Fig:1}(a)], possessing a Rabi frequency $\Omega_R=0.1072 \pm 0.0021$. 
These values stem from fitting  
$\tilde{\Omega}_R(\Delta)=\sqrt{\left(\Omega_R\right)^2+\left(\Delta-\Delta_{+_i}\right)^2}$
to the simulated rf spectra.
This resonance corresponds to the lowest energetically interacting state of a spin-$\uparrow$
impurity with the $^6$Li bath [Fig.~\ref{Fig:2}(a)] verifying the existence of a repulsive polaron in our 1D 
setup.
Further resonances corresponding to higher excited states can be identified as e.g. for   
$\Delta_{+} \approx 3.6$ possessing a much lower Rabi frequency. 
To identify the transition that leads to the occurrence of the above-mentioned quasiparticle peak, 
i.e. $\ket{\Psi_{0}(0)} \leftrightarrow \ket{1}$ schematically illustrated in Fig.~\ref{Fig:2}(a),
we first compute the energy, $E_i(g)$, ($i=\ket{1}$) for this configuration. 
The resulting energy difference, $\Delta E_i(g)=\left[E_i(5)-E_i(0)\right]/n$, 
with $E(0)$ being the energy of the initial state 
and $n$ the order of the transition, 
is the one that matches the location of the observed resonance. 
The corresponding polaronic energy branch shows a monotonic increase for increasing interspecies
repulsion [see the light blue line in Fig.~\ref{Fig:2}(d)],
a behavior that is consistent with the experimental~\cite{Scazza} 
and the theoretical predictions~\cite{Cui,Pilati,Massignan1,Schmidt1} in higher dimensional settings. 
\begin{figure*}[ht]
\includegraphics[width=0.9\textwidth]{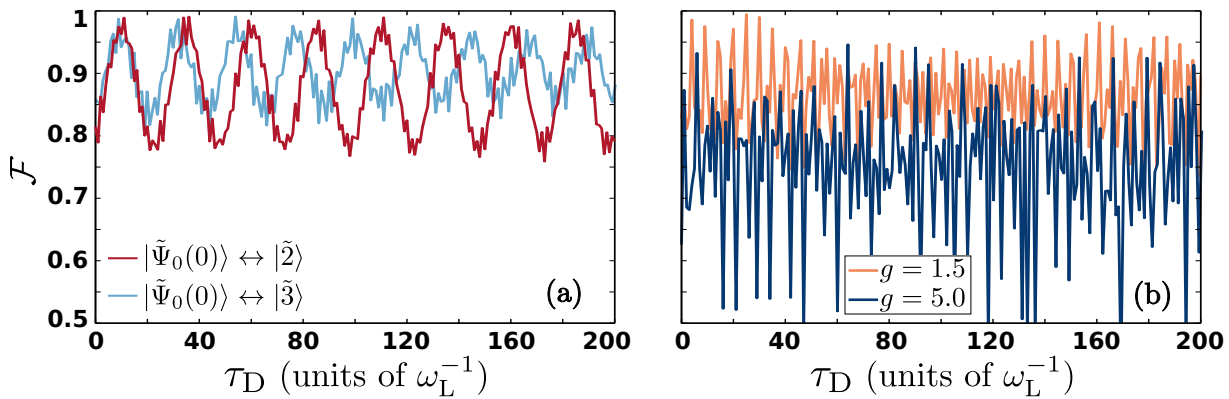}
\caption{(Color online) Simulated spectroscopic signal $\mathcal{F}(\tau_D)$ 
as a function of the dark time $\tau_D$ (see text) showcasing the coherent oscillations of (a) the single  
and (b) the multi-polaron states having (a) strong ($g=5$), 
and (b) strong and weak interspecies repulsions. 
In all cases $N_{L}=5$, $N_K=2$.} 
\label{Fig:3}
\end{figure*}

As a next step we consider a single impurity being initialized in its first excited state
$\ket{\Psi_{1}(0)}$. 
This	 is of importance for the case of $N_K=2$ impurities 
for which more transitions are possible.
In sharp contrast to the $\ket{\Psi_{0}(0)}$ case, two dominant 
polaron peaks appear in the rf spectrum of Fig.~\ref{Fig:1}(b)
centered at $\Delta_{+_{\ket{2}}}=2.152\pm0.001$ ($\Omega_R=0.0899 \pm 0.0012$) and $\Delta_{+_{\ket{3}}}=2.688\pm 0.002$ 
($\Omega_R=0.05072\pm0.022$) respectively. 
These two quasiparticle peaks occur at lower and higher values of $\Delta$ respectively, 
when compared to the $\Delta_{+_{\ket{1}}}$ resonance. 
The width of the resonance centered at $\Delta_{+_{\ket{3}}}$ is significantly sharper compared to the lower-lying one 
as it possesses lower $\Omega_R$. 
The corresponding transitions in this case namely 
$\ket{\Psi_{1}(0)} \leftrightarrow \ket{2}$, and $\ket{\Psi_{1}(0)} \leftrightarrow \ket{3}$ 
are shown in Fig.~\ref{Fig:2}(b).  
The relevant energy branches, $\Delta E_{\ket{2}}(g)$, $\Delta E_{\ket{3}}(g)$, for increasing $g$ are depicted 
in Fig.~\ref{Fig:2}(d). 
It is evident that for $g\leq 1.5$ all the aforementioned resonances
except the transition $\ket{\Psi_{1}(0)} \leftrightarrow \ket{3}$, are overlapping since
$\Delta E_{\ket{1}}(g) \simeq \Delta E_{\ket{2}}(g)$. However, $\Delta E_{\ket{3}}(g)$ possesses a non-zero value 
even for $g\approx0$ as the involved states are already distinct [Fig.~\ref{Fig:2}(b)]. 

To probe the existence of effective interactions between polarons we next consider the case of 
two $^{40}$K impurities immersed in the $^{6}$Li sea. 
Figure~\ref{Fig:1}(c) shows the rf spectrum for $N_K=2$, and $g=5$. 
Here, three narrowly spaced resonances can be observed, see the broad structure centered around $\Delta=2.4$
in Fig.~\ref{Fig:1}(c). 
This broadening together with an overall small upshift  
with respect to the above single impurity cases, 
has been argued to be indicative of the presence of induced interactions between 
the polarons~\cite{Scazza,Cetina} that we will explore below.
The resonances are located at  
$\Delta_{+_{\ket{\tilde{2}}}}=2.195 \approx \Delta E_{\ket{\tilde{2}}}$ ($\Omega_R=0.0836 \pm 0.0004$), 
$\Delta_{+_{\ket{\tilde{1}}}}=2.441 \approx \Delta E_{\ket{\tilde{1}}}$ ($\Omega_R=0.0745 \pm 0.001$), 
and $\Delta_{+_{\ket{\tilde{3}}}}=2.722 \approx \Delta E_{\ket{\tilde{3}}}$ ($\Omega_R=0.0577 \pm 0.0005$) 
respectively.
The relevant transitions are $\ket{{\tilde\Psi}_{0}(0)} \leftrightarrow \ket{\tilde{2}}$, 
and $\ket{\tilde{\Psi}_{0}(0)} \leftrightarrow \ket{\tilde{3}}$ for the outer resonances,
in direct analogy with the ones found in the single impurity case of Fig.~\ref{Fig:2}(b). 
More importantly herein, the central resonance accounts 
not only for a transition $\ket{{\tilde\Psi}_{0}(0)} \leftrightarrow \ket{\tilde{1}}$ 
but it also involves several second-order processes 
namely $\ket{\tilde{\Psi}_{0}(0)} \leftrightarrow \dots \leftrightarrow \ket{\tilde{4}}$ 
and thus corresponds to a multi-polaron state [Fig.~\ref{Fig:2}(c)].
We showcase this, by calculating the probability of finding two particles 
with spin-$\uparrow$~[see the inset in Fig.~\ref{Fig:1}(c)].
It is the appearance of this $\ket{\tilde{4}}$ state that leads to higher-order transitions via the 
virtual occupation of $\ket{\tilde{4}}$ [Fig.~\ref{Fig:2}(c)]. 
The observed upshift of all spectral lines is attributed to the occurrence of this state.
Strikingly enough, the energy of this two-polaron state, $\Delta E_{\ket{\tilde{4}}}$, 
almost coincides with the single polaron one, $\Delta E_{\ket{1}}$, i.e. it exhibits a deviation of $1.9\%$
which is of the same order as the observed upshift [Fig.~\ref{Fig:2}(d)]. 
Note that such a two-polaron resonance is also present for weaker interactions located at $\Delta_{+} \approx 1.175$, 
see Fig.~\ref{Fig:1}(d) and its inset for g=1.5~\cite{comment3}.
Thus, increasing the number of impurities does not significantly 
affect the energy of the polaron or the multi-polaron state formed, 
in accordance with the absence of a significant shift of the corresponding energy in current experimental 
settings~\cite{Scazza}. The origin of the above-mentioned positive shift can be further attributed to the difference 
between the effective and bare mass of the impurities~\cite{Schirotzek,Scazza}, as well as 
to the presence of induced interactions between the polarons \cite{Mora,Yu,Guardian}. 
Therefore the position of the resonance might not be an adequate experimental probe for  
the presence of induced interactions. 
Indeed, the observed energy shift between the energy of the single and two impurites 
is rather small, being of the order of $1.9\%$, and therefore given the current experimental resolution 
it might even not be easily experimentally detectable. 
Instead as we shall demonstrate below the spatial separation of the impurities is the relevant quantity
and can be probed by current state-of-the-art experimental methods.

\subsection{Residue and Lifetime of the Polaron}

To further characterize the polarons we employ their residue, $Z_{fi}$,
which is a measure of the overlap between the dressed polaronic state and
the initial non-interacting one after a single spin flip~\cite{Schmidt3,Massignan3}. 
It is important to note here that in one-dimension the quasiparticle residue acquires a finite value for any finite $N_L$. 
Indeed, the Anderson orthogonality catastrophe occurs only in the thermodynamic limit $N_L\to\infty$ \cite{Anderson_cat,Massignan3} 
rendering the quasiparticle picture ill defined. 
We have used two independent ways for determining $Z_{fi}$. 
Initially, with the aid of Fermi's golden rule, $\Gamma_{i \rightarrow f}= 
2\pi \hbar \left(\Omega^0_R\right)^2 \sum_{f} Z_{fi} \delta\left( 
\omega-\omega_f\right)$,  where $Z_{fi}\equiv|\braket{f|\hat{S}_x|i}|^2$, 
we can deduce that the residue is related to our simulated rf procedure via 
$Z_{fi}=\left(\Omega_R / \Omega^0_R \right)^2 \equiv Z^{\rm{rf}}_{fi}$~\cite{Kohstall,Scazza}. 
For the three polaron peaks identified in Fig.~\ref{Fig:1}(c) the above gives: $Z^{\rm{rf}}_{\ket{\tilde{\Psi}_{0}(0)} 
\ket{\tilde{2}}}=0.5107 \pm 0.0136$, $Z^{\rm{rf}}_{\ket{\tilde{\Psi}_{0}(0)} \ket{\tilde{1}}}=0.7277 \pm 0.0285$, and 
$Z^{\rm{rf}}_{\ket{\tilde{\Psi}_{0}(0)} \ket{\tilde{3}}}=0.1629 \pm 0.0141$. 
Additionally, one can calculate the quasiparticle weight
by invoking its definition. 
The corresponding $Z_{if}$'s are presented in Fig.~\ref{Fig:1}(e) upon varying $g$.
For increasing $g$ $Z_{if}$ decreases being dramatically steeper for the multi-polaron 
state, $Z_{\ket{\tilde{1}} \ket{\tilde{4}}}$, when compared to the single polaron case $Z_{\ket{\tilde{\Psi}_{0}(0)} \ket{\tilde{1}}}$. 
This result supports the observation that polarons consist of well-defined quasiparticles
in the single impurity limit~\cite{Koschorreck}. 
Importantly here, very good agreement
in evaluating $Z_{fi}$ is observed between the two approaches as can be seen
by comparing e.g. at $g=5$  $Z_{\ket{\tilde{\Psi}_{0}(0)} \ket{\tilde{3}}}=0.1627$ shown 
in Fig.~\ref{Fig:2}(e) to $Z^{\rm{rf}}_{\ket{\tilde{\Psi}_{0}(0)} \ket{\tilde{3}}}$. 

The coherence properties of the above-identified polarons can be directly inferred by measuring their lifetime. 
Due to the 1D confinement and due to the fact that in the Hamiltonian of 
Eq.~(\ref{Htot}) incoherent two- and three-body recombination processes are 
ignored~\cite{Massignan3,Massignan1,Schmidt1,Schmidt3,Petrov}, 
only coherent oscillations are expected and indeed observed. 
Figures~\ref{Fig:3}(a) and \ref{Fig:3}(b) summarize our findings for $N_K=2$ both for weak and strong coupling. 
To obtain these lifetimes a two-pulse rf scheme is adopted, mimicking the experimental procedure~\cite{Kohstall},
which is briefly outlined here [see Appendix \ref{rf_implement} for details].
For a specific resonance a $\pi$-pulse is applied transferring the atoms from their initial 
spin-$\downarrow$ to their polaronic spin-$\uparrow$ state. 
Then the particles are left to evolve in the
absence of an rf field, $\Omega^0_R=0$, for a variable (dark) time, $\tau_D$.
After this dark time a second $\pi$-pulse is used driving the impurities 
from the interacting (spin-$\uparrow$ state) to their non-interacting (spin-$\downarrow$) state. 
The signature of this process is the fraction of atoms transferred to the spin-$\downarrow$ state 
during the second pulse divided by the transferred atoms during the first pulse. 
Namely
\begin{equation}
\mathcal{F}(\tau_D)=\frac{\Big[f\left(\frac{\pi}{\Omega_R} \right)-f\left(\frac{2 \pi}{\Omega_R} + \tau_D 
\right)\Big]}{f\left(\frac{\pi}{\Omega_R} \right)}. 
\end{equation}
Note that the presence of excitations as well as higher-order 
transitions, signify the non-adiabatic nature of this procedure.
Thus a phase difference between the distinct polaronic states contributing to the MB wavefunction
is accumulated during the dark time leading in turn to the observed oscillations [Fig.~\ref{Fig:3}].  
Evidently, for single particle transitions a dominant oscillation frequency can be deduced [Fig.~\ref{Fig:3}(a)],
whereas multiple ones occur in the corresponding two-polaron case [Fig.~\ref{Fig:3}(b)] due to the virtual occupation 
of the $\ket{\tilde{4}}$ state. 

\begin{figure*}[ht]
\includegraphics[width=1.0\textwidth]{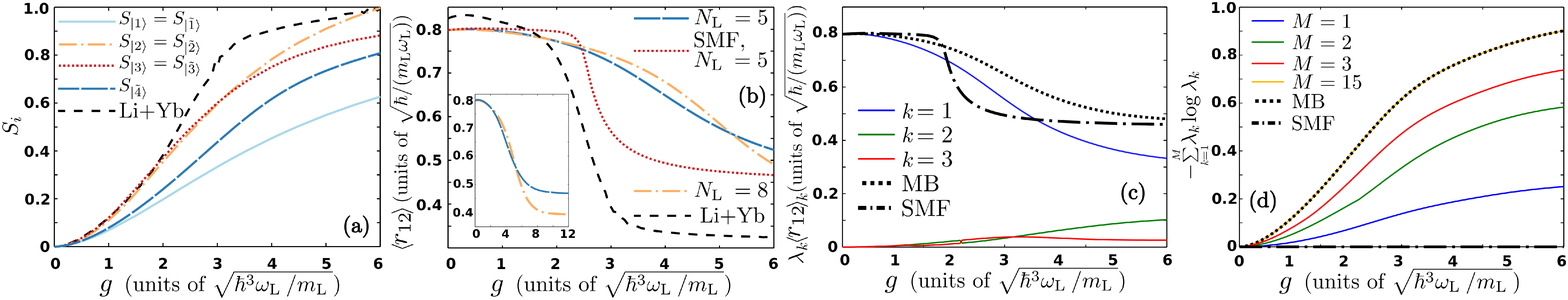}
\caption{(Color online) (a) Von-Neumann entropy, $S_i$, upon varying the interspecies repulsion for all the 
identified, $i=\ket{1},\dots,\ket{\tilde{4}}$, transitions of Fig.~\ref{Fig:2}(c).   
(b) Relative distance, $\langle{r_{12}\rangle}$, between the polarons in the multi-polaron state (see text) dictating the presence
of induced interactions both in the SMF and the MB case (see legend and text). 
The inset illustrates $\langle{r_{12}\rangle}$ within the MB case for larger repulsions. 
(c) Relative distance calculated for the distinct modes of entanglement, $\lambda_k \langle{r_{12}\rangle}_k$ (with $k=1,2,3$), in the 
multi-polaron case for increasing interspecies repulsion $g$. 
(d) Cumulative von-Neumann entropy upon consecutively adding species functions i.e. $M=1, \ldots, 15$ until the MB result is reached. 
In (c), (d) both the MB and the SMF results are illustrated (see legend), while the findings correspond to the case of a $^6$Li-$^{40}$K mixture 
consisting of $N_{L}=5$ and $N_{K}=2$ particles.} 
\label{Fig:4}
\end{figure*}

\section{Entanglement and Induced Interactions} \label{entanglement}

To unravel the entangled nature of both the single and the two-polaron states we next invoke the 
von-Neumann entropy~\cite{Horodeki}. 
It is important to stress here that the polaronic states refer to stationary states of the binary mixture consisting 
of bath and spin-$\uparrow$ atoms. 
For this binary system ($\sigma=L,\uparrow$)  the von-Neumann entropy reads 
$S_{i}=-\rm{Tr}_{\sigma} \left[\rho^{\sigma} \log\left( \rho^{\sigma} \right)\right]$, with
$\rho^{\sigma}=-\rm{Tr}_{\sigma'} \left[\ket{\Psi} \bra{\Psi} \right]$ being the species density matrix. 
In Fig.~\ref{Fig:4}(a) $S_i$ is shown for all of the above transitions, namely $i=\ket{1},\dots,\ket{\tilde{4}}$, as a function of the 
coupling strength. In all cases a monotonic increase of $S_i$ is observed when entering deeper into the repulsive regime.
Strikingly enough, the entropy is found to be significant not only for the two-polaron state
but also for the single polaron ones suggesting that these quasiparticles are in general entangled. 
However notice the deviation between the single ($S_{\ket{1}}$) and the two-polaron state ($S_{\ket{\tilde{4}}}$) 
which is of the order of $20\%$ for large repulsions. 

Turning our attention to the two-polaron state, we aim to reveal the presence of induced interactions. 
As discussed above one cannot necessarily infer about the latter by solely considering the energies. 
Therefore we employ the relative distance between the two $^{40}$K impurities that constitute 
the multi-polaron state for variable $g$. 
The relative distance reads
\begin{equation}
\begin{split}
&\langle{r_{12}\rangle}=\\ &\frac{\int dx_1 dx_2 |x_1-x_2| \braket{\Psi|\hat{\Psi}^{\dagger}_{\uparrow} (x_1) 
\hat{\Psi}^{\dagger}_{\uparrow} (x_2) \hat{\Psi}_{\uparrow} 
(x_2) \hat{\Psi}_{\uparrow} (x_1)|\Psi}}{
\braket{\Psi|\hat{N}_{\uparrow} \left(\hat{N}_{\uparrow} -1\right)|\Psi}},    
\end{split}
\end{equation} 
where $\hat{\Psi}_{\uparrow}(x_1)$ denotes the fermionic field operator that annihilates a fermion at position $x_1$. 
$\hat{N}_{\uparrow}$ is the number operator that measures the number of fermions residing in the spin-$\uparrow$ state. 
Such a quantity can be directly probed experimentally by performing {\it in-situ} spin-resolved single-shot measurements
on the $\uparrow$-state of $^{40}$K~\cite{Jochim2}. 
Each image offers an estimate of the relative distance between the polarons provided that the 
position uncertainty is relatively low~\cite{Jochim2}. 
Then $\braket{r_{12}}$ is obtained by averaging over several such images. 
Evidently [see Fig.~\ref{Fig:4}(b)] stronger repulsions result in a significant decrease 
of $\langle{r_{12}\rangle}$ that drops to almost half of its initial value for $g\geq 5$. 
In this way, $\braket{r_{12}}$ clearly captures the manifestation of 
attractive induced interactions present in the system saturating for even larger g [see the inset in Fig.~\ref{Fig:4}(b)]. 
As shown in Fig. \ref{Fig:4} (b) this behaviour of $\braket{r_{12}}$ holds equally for larger particle numbers 
of the bath, i.e. $N_{L}=8$, and different atomic species, e.g. a $^{6}$Li-$^{173}$Yb mixture 
possessing $\omega_{Yb}=0.125~\omega_{L}$~\cite{Yb1,Yb2}. 
This indicates that $\langle{r_{12}\rangle}$ 
captures the presence of induced interactions independently of the specifics of the atomic mixture.
It becomes also apparent that heavier impurities lead to even stronger attraction 
emerging from drastically smaller interactions. 
Most importantly, by calculating $\langle{r_{12}\rangle}$ in the non-entangled SMF approximation [see also Eq. (\ref{eq:SMF})], 
it can be clearly deduced that its shape, being much smoother in the MB approach [Fig.~\ref{Fig:4}(b)], 
bears information regarding the generation of entanglement [see also our discussion below]. 
Recall that in this latter SMF case the wavefunction ansatz assumes the form 
$\ket{\Psi}_{SMF}=\ket{\tilde{\Psi}^{L}} \otimes \ket{\tilde{\Psi}^{\uparrow}}$, which is the most general ansatz that excludes entanglement but 
includes intraspecies correlations.  

Therefore, $\langle{r_{12}\rangle}$ is indicative of the generation of entanglement, as dictated by the growth 
rate of $S_i$ for varying $g$, in MB systems. 
Indeed, the relation between the generation of entanglement and the $\langle{r_{12}\rangle}$ can be understood as follows. 
In order to connect the relative distance with the generation of entanglement 
we must first recall that the system under consideration is a bipartite composite system
whose MB wavefunction, $\ket{\Psi}_{MB}$, can be expressed in terms of the truncated Schmidt decomposition of rank $M$ as
\begin{eqnarray}
\ket{\Psi}_{MB}=\sum^M_{k=1} \sqrt{\lambda_k} \ket{\tilde{\Psi}^{L}_k } \ket{\tilde{\Psi}^{\uparrow}_{k}}.
\label{schmidt}
\end{eqnarray} 
Here $\ket{\tilde{\Psi}^{L}_k }$ and $\ket{\tilde{\Psi}^{\uparrow}_{k}}$ denote the species wavefunction 
of the bath and the impurity respectively. 
The weights $\lambda_k$ in decreasing order are referred to as the natural occupations of the $k$-th species function, 
and $\sqrt{\lambda_k } \ket{\tilde{\Psi}^{L}_k } \ket{\tilde{\Psi}^{\uparrow}_{k} }$ denotes the $k$-th mode of entanglement.
Then the expectation value $\langle{r_{12}\rangle}$ 
in terms of the Schmidt coefficients $\lambda_k$ reads 
\begin{eqnarray}
\begin{split}
\langle{r_{12}\rangle}&= \sum^M_{k=1} \lambda_k \int dx_1 dx_2~|x_1-x_2| \times \\
&\frac{\braket{\hat{\Psi}^{\uparrow}_{k}  
|\hat{\Psi}^{\dagger}_{\uparrow} (x_1) 
\hat{\Psi}^{\dagger}_{\uparrow} (x_2) \hat{\Psi}_{\uparrow} 
(x_2) \hat{\Psi}_{\uparrow} (x_1)|\hat{\Psi}^{\uparrow}_{k}}}{
\braket{\Psi|\hat{N}_{\uparrow} \left(\hat{N}_{\uparrow} -1\right)|\Psi}} \\ 
&\equiv \sum^M_{k=1} \lambda_k (t) \langle{r_{12}\rangle}_k.   
\end{split}
\end{eqnarray}
It becomes apparent by the above expression that the interplay of two different quantities has to be taken into account
in order to extract the dominant contribution that leads to the final shape of $\langle{r_{12}\rangle}$ when including all the 
relevant correlations. 
Namely the Schmidt weights, $\lambda_k$, 
and the two-body correlator $\langle{r_{12}\rangle}_k$ of the $k$-th mode of entanglement. 
In Fig.~\ref{Fig:4} (c) $\lambda_k\langle{r_{12}\rangle}_k$ 
is illustrated for each of the first three individual species functions $k=1,2,3$, and for the case of a $^6$Li-$^{40}$K mixture
consisting of $N_L=5, N_K=2$ fermions. 
Also in the same figure we have included the corresponding full MB result 
depicted with the dashed dotted black line, as well as the relevant outcome 
in the non-entangled SMF case [see the dashed black line in Fig.~\ref{Fig:4} (c)]. 
Notice the abrupt decrease of $\langle{r_{12}\rangle}$ in the SMF case when compared to the 
much smoother decay observed in the presence of entanglement. It is exactly this comparison of the MB outcome to 
the SMF one which reveals that the relative distance itself via its shape bears information regarding the generation of entanglement  
in the system.   
Additionally, as can be clearly deduced from this figure the dominant contribution 
to the final shape of $\langle{r_{12}\rangle}$ stems from {\bf $\lambda_1\langle{r_{12}\rangle}_1$} 
[see solid blue line in Fig.~\ref{Fig:4} (c)] which corresponds to the first mode of entanglement. 
It is important to note here, that the form of this dominant mode, $\lambda_1\langle{r_{12}\rangle}_1$, in the MB case is greatly altered
when compared to the non-entangled, $\langle{r_{12}\rangle}_{SMF}$, case.
Therefore it becomes apparent that besides this dominant contribution 
also higher order modes of entanglement weighted by $\lambda_2, \lambda_3, \ldots $ 
are significant in retrieving the MB outcome indicating the strongly entangled nature of the system. 

Turning to the von-Neumann entropy recall that the latter can be written in terms of the Schmidt coefficients as follows: 
$S_{M}=-\sum^M_{k=1} \lambda_k \log \lambda_k$. The corresponding $S_{\ket{\tilde{4}}}\equiv S_{M}$ upon consecutively adding higher order contributions
is shown in Fig.~\ref{Fig:4} (d). 
Indeed inspecting Fig.~\ref{Fig:4} (d) it becomes evident that in order to retrieve the full MB result the higher-lying Schmidt 
coefficients, namely $k>1$, are the ones that predominantly contribute to the final shape of $S_{M}$. 
This result is in sharp contrast to the behavior of the relative distance which is mainly determined by the first mode of entanglement characterized by the 
leading order Schmidt coefficient, namely the $\lambda_1$. 
Notice also that in the same figure we have included the corresponding SMF result 
[see the dashed black line in Fig.~\ref{Fig:4} (d)]
just to showcase that in this case the von-Neumann entropy is zero due to the absence of entanglement. 

It becomes evident by the above discussion that both the von-Neumann entropy and the relative distance dictate the 
generation of entanglement in the MB system but by taking into account different contributions. 
Additionally, since both quantities are given in terms of the Schmidt coefficients being subject to the constraint 
$1-\lambda_1=\sum^M_{i=2} \lambda_i$, when $\Delta r_{12} (g)=\langle{r_{12}\rangle}_{MB}-\langle{r_{12}\rangle}_{SMF}$ 
is finite then also $\Delta S_{M} (g)=S_{M_{MB}}-S_{{SMF}}$ is finite. 
Moreover, $S_{M} (g)$ is used to showcase that polarons are indeed entangled with their environment. 
However, since $S_{M}$ cannot be measured experimentally, one can infer about the 
generation of entanglement in the MB system via the shape of $\langle{r_{12}\rangle}$ 
which can be probed via in-situ spin-resolved 
single-shot measurements that are nowdays available~\cite{Jochim2}.
It is also worth commenting at this point that the above results can be generalized to any type of mixture not necessarily 
a fermionic one.

\section{Conclusions} \label{conclusions}

We have investigated the existence and emergent properties of single and multiple repulsive polarons in 
1D harmonically confined fermionic mixtures both for weak and strong interspecies interactions. 
In particular, we have simulated the corresponding experimental process of radiofrequency spectroscopy using different fermionic mixtures 
consisting of a single or two impurities coupled to a few-body Fermi sea. 
Analysing the obtained radiofrequency excitation spectrum it is indeed shown that these impurities accumulate polaronic properties. 
Most importantly, we identify all dominant microscopic mechanisms that lead to the polaron formation. 
We verify that by increasing the number of impurities immersed in a bath with fixed particle number both 
single and multi-polaron states occur independently of the interaction strength. 
The corresponding polaronic states are characterized by extracting their residua and lifetimes. 
We find that the residue exhibits a decreasing tendency for increasing interspecies interaction strengths. 
This decrease is found to be much more prominent for a multi-polaron than a single polaron state. 
On the other hand the spectroscopic signal shows an oscillatory behavior with variable dark time indicating 
the longevity of the polarons. 

Turning to the induced interactions between the polarons we show that their presence is first dictated 
by a positive resonance shift in the radiofrequency spectrum accompanied by a consequent spectral broadening. 
This latter finding is in accordance with the recent experimental observations in three-dimensional setups. 
However, the above-mentioned shift possesses a small amplitude, being of the order of $2\%$. 
This implies that in order to infer about the presence of induced interactions an alternative measure is needed. 
Our alternative measure for probing the presence of induced interactions is the relative distance between the polarons. 
It can be experimentally probed via {\it in-situ} spin-resolved single-shot measurements. 
Attractive induced interactions are indeed captured by this quantity and shown to persist upon enlarging the fermionic 
sea or considering different fermionic species. 
The shape of the relative distance for increasing interspecies interactions is found to be also indicative of the 
presence of entanglement in the MB system. 
To quantify the degree of entanglement between the impurity and the bath we resort to the von-Neumann entropy 
which acquires finite values and in particular increases for larger interactions. 
The degree of entanglement is found to be crucial for the case of a single and for two impurities, 
being larger in the latter case. 

Our investigation of strongly correlated 1D repulsive fermi polarons and multi-polaron states opens up the possibility 
of further studies of quantum impurities in lower dimensional settings. 
In particular a straightforward extension of our results would be to consider bosonic or fermionic impurities of the same or higher 
concentration in a bosonic bath and study the consequent formation of quasiparticles.  
An imperative prospect would be to examine the existence and properties of such quasiparticle states in the 
1D to the 3D crossover, an investigation that calls for further experimental studies. 
Another interesting direction is to unravel the few-to-many-body crossover regarding the size of the bath in order to reveal 
its impact on the emergent polaronic properties. 
Certainly the study of dressed impurities in the strongly interacting regime where the polaron picture is expected to break down 
is an intriguing prespective.

\appendix

\section {Details of the Reverse Radiofrequency Spectroscopy} \label{rf_implement}

The purpose of this section is to elaborate on the model that allows for the
simulation of radiofrequency (rf) spectroscopy \cite{Kohstall,Scazza}.  The
latter has been employed in the main text for the identification of the
polaronic resonances and the subsequent characterization of their coherence
properties.

In our case few ${}^{40}$K atoms are immersed in an environment consisting of
${}^6$Li atoms close to an interspecies magnetic Feshbach resonance (FR)
\cite{Chin}. Such resonances occur at magnetic fields of the order of $100$ G
\cite{Naik,Feshbach2,Tiecke}, where the ground state of ${}^{40}$K
atoms, $|{}^2S_{1/2};F=\frac{9}{2}\rangle$, experiences a sizeable quadratic
Zeeman shift \cite{Parameters}. This Zeeman shift allows us to address
selectively the distinct $m_F$ transitions provided that the employed intensity
of the rf pulse results in a Rabi frequency $\Omega_R$ much smaller than the
Zeeman splitting of the involved hyperfine levels. In this work we consider
two such hyperfine levels of ${}^{40}$K denoted as $\ket{\uparrow}$, 
$\ket{\downarrow}$ that can be identified and resonantly coupled for a rf
photon frequency $\nu_{0}$, corresponding to the
Zeeman splitting between the two levels, in the absence of a ${}^6$Li bath. In
such a case, it suffices to treat the ${}^{40}$K atoms as two-level systems.  As
the atoms are confined within a harmonic potential each of the hyperfine levels
is further divided into states of different atomic motion. The average spacing
between these sublevels corresponds to the harmonic trap frequency, $\omega_K$
being of the order of kHz in typical few-atom experiments
\cite{Heidelberg1,Brouzos,Heidelberg3}. In the vicinity of a Feshbach
resonance the energy of these sublevels strongly depends on the interspecies
interaction strength $g$ between the ${}^{40}$K atoms in the
resonantly-interacting hyperfine state and the ${}^{6}$Li environment.
Accordingly the energy of each motional state shifts by
$\Delta_{+}(g)$, from the corresponding
non-interacting one.  In few-atom experiments this shift is of the order of the
trapping frequency ($\sim$ kHz).

\begin{figure*}[ht]
    \centering
    \includegraphics[width=0.8\textwidth]{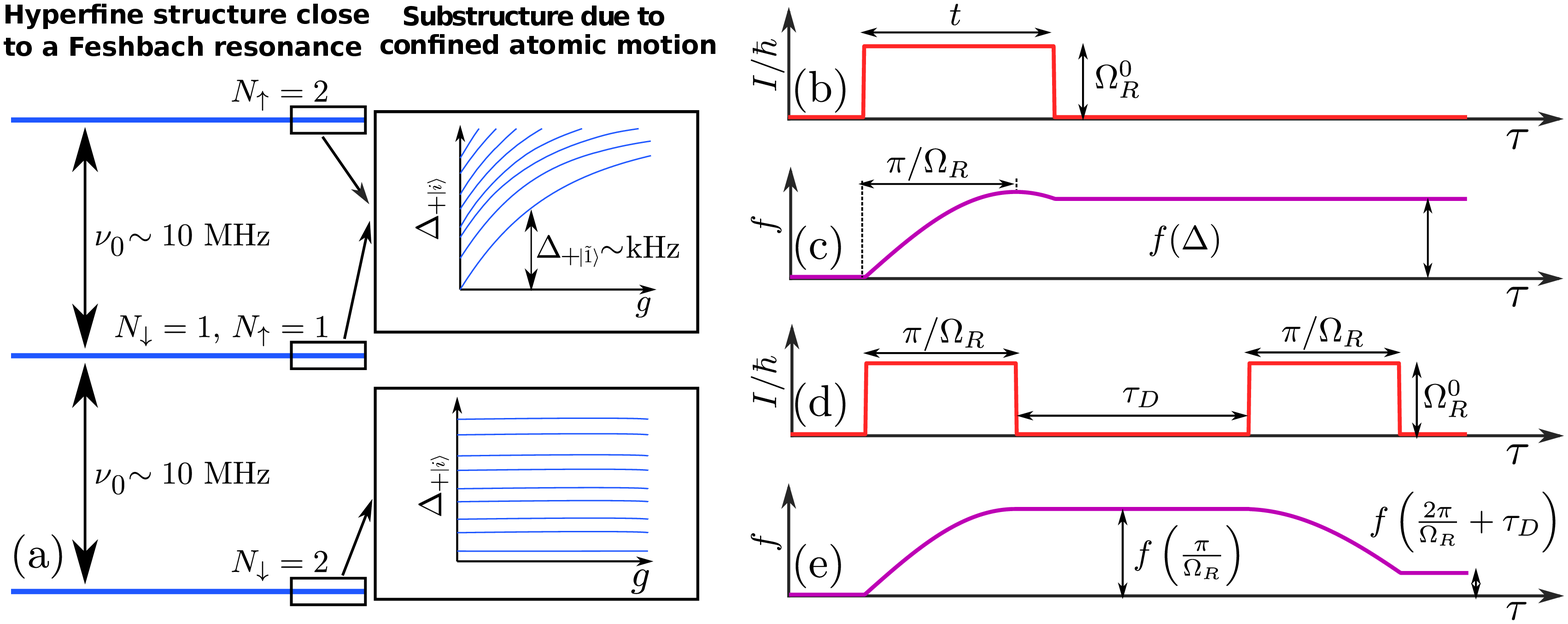}
    \caption{(a) Schematic representation of the involved rf levels for $N_K=2$ ${}^{40}$K atoms coupled to
    a ${}^{6}$Li bath near an interspecies FR at $B\sim 100$G. (b) Intensity $I$ of the employed rf pulse and
    (c) expected time-evolution of the excited fraction $f$ for the detection of polaronic resonances. (d), (e) same as (b), (c) respectively but for
the protocol that probes the coherence properties of the polaronic branch.}
    \label{fig:SM1}
\end{figure*}

Figure~\ref{fig:SM1} (a) schematically demonstrates the rf spectral lines in the case of
$N_K=2$, including resonant interactions between the $\ket{\uparrow}$
particles and the ${}^6$Li environment. Three well-separated energy level
manifolds occur corresponding to the different configurations of $N_{\uparrow}$
and $N_{\downarrow}$, with $N_{\uparrow} + N_{\downarrow}=N_K$, separated by
the Zeeman splitting $\nu_{0}$. Each of these manifolds
exhibits a substructure of different energy levels of atomic motion. For the
configuration $N_{\uparrow}=0$ and $N_{\downarrow}=2$ this substructure is
interaction-independent in sharp contrast to the $N_{\uparrow}=1$,
$N_{\downarrow}=1$ and $N_{\uparrow}=2$, $N_{\downarrow}=0$ configurations as
the $\ket{\downarrow}$ atoms do not interact with neither the $\ket{\uparrow}$ 
${}^{40}$K or the ${}^{6}$Li atoms. Reverse rf spectroscopy can be
employed to identify these interaction energy shifts provided that the Rabi
frequency satisfies $\Omega_R < \Delta_{+}\sim$
kHz. This allows us to invoke the rotating wave approximation as $\Omega_R
\sim$ kHz $\ll \nu_{0} \sim 10$ MHz. Employing this
approximation the Hamiltonian for the internal state of the ${}^{40}$K atoms,
in the interaction picture of the $\uparrow \downarrow$ transition, reads
$\hat{H}_S=-\frac{\hbar \Delta}{2} \hat{S}_z+\frac{\hbar \Omega_R^0}{2} \hat{S}_x$. The
latter is exactly the form employed in the main text.  $\Omega_R^0$ and
$\Delta$ refer to the Rabi frequency and detuning with respect to the resonance
of the $\uparrow \downarrow$ transition at $g=0$. We remark that the
$\ket{\uparrow}$ and $\ket{\downarrow}$ states in the Schr{\" o}dinger and
interaction pictures are equivalent, so our conclusions are invariant under
this frame transformation.

One-dimensional (1D) ensembles offer a clean realization of few-body rf
spectroscopy as the existing bound state of a Feshbach molecule possesses a binding energy of the order of 
$\epsilon_b=-2~\hbar \omega_{\perp}$~\cite{Bloch} at the confinement-induced resonance, i.e. $g_{1D}\rightarrow\infty$.
Since current state-of-the-art few-body 1D experiments have been consistently described by pure 1D models~\cite{Heidelberg3,Frank}
the effect of the bound state for repulsive interactions sufficiently below the $g_{1D}\rightarrow\infty$ regime is 
negligible. Indeed in order to ensure the validity of the 1D description 
$\omega_{\perp}\gg N \omega_{L}$ must hold, where $N$ denotes the total particle number. 
In the worst case scenario considered in the main text, namely that of $N_{L}=8$,
and $N_{K}=2$, and in particular when $\omega_{\perp}= N \omega_{L}$ then $\epsilon_b \approx -20~\hbar~\omega_{L}$. 
However, the detuning parameter, $\Delta$,   
used herein is maximally $4~\hbar~\omega_{L}$. 
Therefore it lies far below the above threshold of $\abs{\epsilon_b}$.
The same line of argumentation holds for the corresponding binding energy at the magnetic FR 
where $\epsilon_b =-0.606~\hbar~\omega_{\perp}$~\cite{Bloch}. 
Additionally, few-body systems involve
low-densities thus drastically reducing the incoherent processes such as two- and
three-body recombination and resulting in increased coherence times. The above
allows us to assume a coherent evolution during the simulated experimental
sequence.

To identify the resonances corresponding to polaronic states we employ the rf
pulse shape depicted in Fig.~\ref{fig:SM1} (b). The system is initialized in the
non-interacting ground state where the ${}^{40}$K atoms are spin-polarized in
their $\ket{\downarrow}$ state and a rectangular pulse of frequency $\nu$,
and detuning $\Delta$ is
employed. This pulse is further characterized by an exposure time $t$ and a
Rabi-frequency $\Omega_R^{0}$. Different realizations utilize different
detunings $\Delta$ but the same $t$ and $\Omega_R^{0}$. In the duration of the
pulse the system undergoes Rabi-oscillations [see Fig.~\ref{fig:SM1}(c)] whenever the
detuning $\Delta$ is close to a resonance $\Delta \approx \Delta_{+}$. The
employed spectroscopic signal is the fraction of atoms transferred to the 
$\ket{\uparrow}$ hyperfine state, namely $f(\Delta,t)=\frac{\langle N_{\uparrow}
\rangle}{N_K}$. We remark that different pulse shapes have been simulated e.g.
Gaussian-shaped pulses, which do not alter the presented results. To infer
about the coherence properties of the polaronic states we employ a Ramsey like
process, see Fig.~\ref{fig:SM1} (d). Initially, we prepare the system in the same
non-interacting ground state as in the previously examined protocol and apply a
rectangular $\pi$-pulse on a polaronic resonance. This sequence transfers the
atoms from the ground state to the polaronic state in an efficient manner. Then
we let the system evolve in the absence of rf fields, $\Omega_R^{0}=0$, for a
dark time, $\tau_D$. Finally, we apply a second $\pi$-pulse identical to the
first one to transfer the atoms from the polaronic to the initial ground state.
The spectroscopic signal is the fraction of atoms that have been excited to the
polaronic branch by the first pulse and subsequently deexcited by the second
one divided by the total number of excited atoms,
$\mathcal{F}(\tau_D)=\frac{f(\pi/\Omega_R)-f(2\pi/\Omega_R+\tau_D)}{f(\pi/\Omega_R)}$,
see also Fig.~\ref{fig:SM1} (e).
\begin{figure*}[ht]
    \centering
    \includegraphics[width=0.8\textwidth]{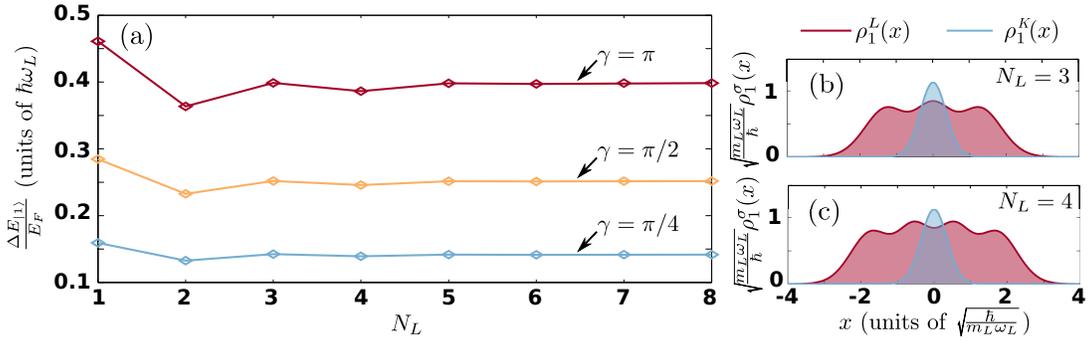}
    \caption{(a) Energy of the polaron for increasing particle number of the bath $N_L$ at various values of the Lieb-Liniger parameter $\gamma$ (see legend). 
     One-body density of the ground state of each species (see legend) at the noninteracting limit, $g=0$, for the case of (b) $N_L=3$, $N_K=1$ and (c) $N_L=4$, $N_K=1$.}
    \label{fig:SM2}
\end{figure*}

\section{Effective Range Corrections} \label{sec:1d}

Below we briefly discuss the applicability of the Hamiltonian employed in the current work (see Eq.(1) in the main text). 
Notice that this model Hamiltonian assumes that contact interactions dominate the dynamics, ignoring effective range corrections. 
It is well-known that a $^6$Li-$^{40}$K mixture features 
narrow FR~\cite{Naik} 
with the broader ones being at $114~\rm{G}$~\cite{Tiecke} 
and $155~\rm{G}$~\cite{Kohstall} magnetic field respectively.
Among these two resonances the former has been suggested as the most promising and at the same time experimentally 
feasible that can be used to reach the universal regime being $s$-wave dominated and satisfying the condition 
$k_F R^*\ll 1$~\cite{Tiecke}. 
Here, $k_F=\sqrt{\frac{2mN\omega_{L}}{\hbar}}$ is the Fermi momentum where $m$, $N$ is the mass and particle number of the relevant component while $R^*$ is the range 
parameter. 
In contrast, the latter FR which is also the narrower of the two,
suffers from effective range corrections that in turn alter the physics of polarons~\cite{Kohstall}
resulting in enhanced lifetimes of these repulsive states. 
In order to showcase that the model Hamiltonian used herein accurately describes the dynamics 
of repulsive fermi polarons below we provide estimates of the effective range parameter $k_F R^*$ 
for the narrower FR at $155~\rm{G}$, and for all the cases investigated in the main text.
Our results are summarized in Table~\ref{table}. 
In particular, in order to calculate the effective range correction $k_F R^*$ 
for the different cases studied in this work, 
we use as a range parameter $R^*=2700 \times 5.29 \times 10^{-11}~\rm{m}$~\cite{Kohstall},
and as a characteristic axial trapping frequency  
$\omega_{\parallel}\equiv\omega_{L}=2\pi\times 75~\rm{Hz}$~\cite{Jochim2}. 
Note also that $m_{L}=6\times1.66\times 10^{-27}~\rm{kg}$ and $m_K=40/6~m_{L}$. 
As it can be clearly seen in all cases of interest here [see the boldface values in Table~\ref{table}], 
$k_F R^*$ is sufficiently smaller than unity. 
The latter verifies the applicability of the model used and thus the universal behavior, by means of a negligible $R^*$, 
of the physics addressed herein. 
Finally, we remark that for the second fermionic mixture considered in this work, 
namely the $^6$Li-$^{173}$Yb one, it is predicted that such a mixture features broad FRs and thus the model Hamiltonian 
used again accurately describes the polaron dynamics~\cite{Yb1}. 
   
\begin{table}\centering
\begin{tabular}{|c||c|c|}
 \hline
 \multicolumn{3}{|c|}{{\bf Effective range}} \\
 \hline \hline 
Number of particles &~~~$(k_F R^*)_{{^6}Li}$~~~ & ~~~ $(k_F R^*)_{^{40}K}$~~~\\
 \hline \hline
 $N$=1  & $0.0426$ & $\textbf{0.0852}$ \\
 $N$=2 & $0.0603$ & $\textbf{0.1206}$ \\
  $N$=5  & $\textbf{0.0953}$ & $0.1906$  \\
 $N$=8  & $\textbf{0.1205}$ & $0.241$ \\
 \hline 
\end{tabular}
\caption{Effective range parameter, $k_F R^*$, calculated for a $^6$Li-$^{40}$K mixture 
showcasing the validity of the single-channel 1D model Hamiltonian used in the main text.  
The experimental axial trapping frequency is $\omega_{\parallel}\equiv\omega_{L}=2\pi\times 75 \rm{Hz}$~\cite{Jochim2},
and the range parameter at resonance reads 
$R^*=2700\times 5.29\times 10^{-11} \rm{m}$~\cite{Kohstall}. 
Note also that $m_{L}=6\times1.66\times 10^{-27}~\rm{kg}$ and $m_K=40/6~m_{L}$.} 
\label{table}
\end{table}

\begin{figure*}[ht]
    \centering
    \includegraphics[width=0.8\textwidth]{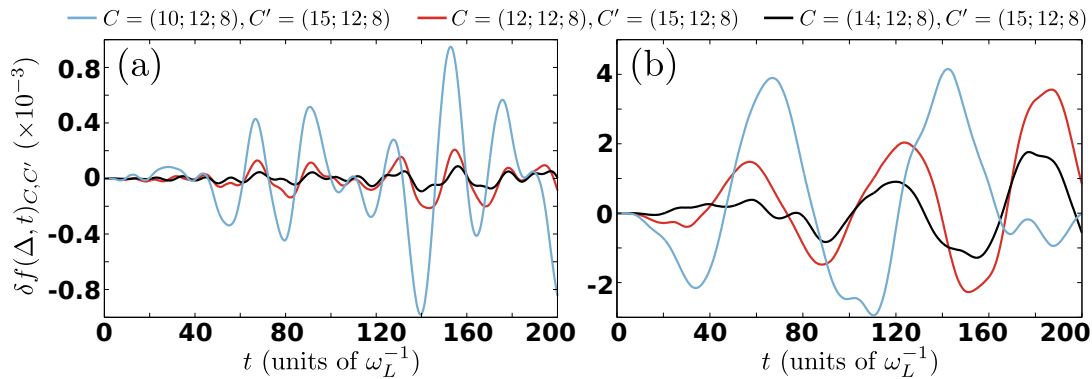}
    \caption{Relative difference of the spectroscopic signal $f(\Delta,t)$ between various orbital configurations 
    $C$ and $C'$ (see legend) with respect to the pulse time $t$. 
    The corresponding rf-detuning is (a) $\Delta=2.2$ and (b) $\Delta=2.5$. 
    The system consists of $N_L=5$ and $N_K=2$ atoms, while the interspecies interaction strength is $g=5$.} 
    \label{Fig:convergence_figure}
\end{figure*}

\section{Polaron Energy versus the particle number of the bath}\label{polaron_energy} 

Let us investigate the behavior of the polaron energy while approaching the MB limit by increasing the number of the bath 
particles $N_L$. 
It is known \cite{Brouzos} that the polaron energy scales proportionally to the square root of the bath particle number, i.e. 
$\Delta E_{\ket{1}}(g)=E_{\ket{1}}(g)-E_{\ket{\Psi_0(0)}}(g)\propto \sqrt{N_L}$. 
Here, $E_{\ket{1}}(g)$ [$E_{\ket{\Psi_0(0)}}$] denotes the ground state energy of the interacting [noninteracting] ($N_L+1$)-body system. 
In order to obtain a non-divergent polaron energy $\Delta E_{\ket{1}}(g)$ for $N_L\to\infty$ we rescale it with the corresponding Fermi 
energy $E_F=N_L\hbar\omega_L$. 
Note that in the presence of a harmonic trap $E_F$ refers to the energy of the energetically lowest unoccupied single-particle eigenstate \cite{Brouzos}. 
Therefore, the rescaled polaron energy is proportional to $\Delta E_{\ket{1}}/E_F \propto 1/\sqrt{N_L}$. 
Moreover, in order to obtain a dimensionless interaction parameter that scales similarly to the polaron energy with respect to $N_L$ we define the 
so-called Lieb-Liniger parameter $\gamma=\frac{\pi m g}{\hbar^2 k_F}$, where $k_F=\sqrt{2m_L E_F}/\hbar$ is the Fermi momentum. 
The interaction interval used in the main text is $0<g<5$ which corresponds to $0<\gamma<\frac{3\pi}{2}$. 

To provide some representative examples of the convergence of $\Delta E_{\ket{1}}/E_F$ for increasing $N_L$ and fixed $\gamma$ we choose the values 
$\gamma=\frac{\pi}{4},\frac{\pi}{2},\pi$, see Fig. \ref{fig:SM2} (a). 
As it can be seen, for $N_L\geq5$ the polaron energy exhibits a saturated behavior. 
The latter observation essentially indicates that the particle number of the bath $N_L=5$ captures adequately the behavior of 
the Fermi polaron at the MB level. 
On the other hand, for smaller particle numbers i.e. $N_L\leq4$ we observe that $\Delta E_{\ket{1}}/E_F$ depends strongly on $N_L$. 
This effect can be attributed to the behavior of the one-body density of the bath which exhibits a local maximum (minimum) in the vicinity of 
$x=0$ for an odd (even) particle number $N_L$. 
Since the impurity is localized around $x=0$ the scaled interaction energy is larger for an odd compared to an even particle number of the bath. 
To provide a concrete example in Figs. \ref{fig:SM2} (b), (c) we demonstrate the ground state one-body densities of each species at the noninteracting limit for the 
systems $N_L=3$, $N_K=1$ and $N_L=4$, $N_L=1$ respectively. 
We observe that in the case of $N_L=3$, $N_K=1$ the densities of the ${}^{6}$Li and ${}^{40}$K possess a larger spatial overlap 
compared to the case of $N_L=4$, $N_K=1$. 
As a consequence the corresponding interaction energy between the species is larger for $N_L=3$ than the $N_L=4$ system. 
In turn, this explains the larger rescaled energy of the polaron in the case of an odd than an even particle number of the bath.

\section{Remarks on The Many-Body numerical method: ML-MCTDHX} \label{sec:numerics1}

To address the MB dynamics during rf spectroscopy we rely on the Multi-Layer
Multi-Configuration Time-Dependent Hartree method for Atomic Mixtures
\cite{moulosx} (ML-MCTDHX). The main distinctive features of the employed method
are outlined below.  First, within ML-MCTDHX the total MB wavefunction is
expanded with respect to a time-dependent and variationally optimized
MB basis.  This allows us to achieve convergence by employing a
drastically reduced number of time-dependent basis states compared to methods relying on
a time-independent basis.  Second, the symmetry of the atomic species being either
bosonic or fermionic is explicitly employed by considering the expansion of the
MB wavefunction in terms of the number-states spanned by the underlying time-dependent basis. 
Finally, the multi-layer ansatz for the total wavefunction is based on
a coarse-graining cascade, where strongly correlated degrees of freedom are
grouped together and treated as subsystems mutually coupling to each other. The
latter enables us to tailor the employed MB wavefunction ansatz according to
the specific intra- and inter-species correlation patterns emanating in
different setups. The latter renders ML-MCTDHX a versatile tool for simulating the dynamics
of multispecies systems.  In particular this work employs a reduction of the
ML-MCTDHX method for mixtures of two fermionic species one of which possesses an
additional spin-$1/2$ degree of freedom.

For our implementation we have used a harmonic oscillator DVR, resulting
after a unitary transformation of the commonly employed basis of harmonic
oscillator eigenfunctions, as a primitive basis for the spatial part of the SPFs. 
To study the dynamics of the spinor system we propagate the wavefunction of Eq.~(\ref{eq:tot_wfn}) by utilizing the
appropriate Hamiltonian within the ML-MCTDHX equations of motion.  

To infer about convergence we demand that all the observables of
interest ($f$, $\mathcal{F}$) do not change within a given relative accuracy (see also below). 
In order to achieve the above criterion we increase the DVR basis states, $\mathcal{M}$,
as well as the number of species wavefunctions, $M$, and SPFs
$m^{\sigma}$ (with $\sigma=A,B$ denoting each of the species).
More specifically,  
for the two different mixtures presented in the main text namely 
the $^6$Li-$^{40}$K and the $^6$Li-$^{173}$Yb mixture the number of grid points used are 
$\mathcal{M}=80$ and $\mathcal{M}=150$ respectively. 
Additionally, for the cases investigated in the main text  
i.e. $N_L=5$ and $N_K=1$, 
$N_L=5$ and $N_K=2$, $N_L=8$ and $N_K=2$, and $N_L=5$ and $N_{Yb}=2$, 
the corresponding configurations satisfying the aforementioned convergence criterion are 
$C=(6;10;6)$, $C=(15;12;8)$, $C=\left(14;14;8 \right)$ and $C=(15;10;10)$ respectively. 
The orbital configuration $C$ follows the notation $C=(M;m^A;m^B)$. 
It is important to note here that e.g. for the case of $N_L=5$ and $N_K=1$, with $C=(6;10;6)$
the truncated Hilbert space 
for the corresponding rf simulation involves $2864$ coefficients, 
while for an exact diagonalization treatment it would require 
the inclusion of $1.9232\times 10^9$ coefficients rendering the latter simulation infeasible. 
The same result also holds for all the cases explored in the main text. E.g. for $N_L=5$ and $N_K=2$ with $C=(15;12;8)$ the inclusion 
of $14.125$ coefficients is needed within the ML-MCTDHX approach, 
while the number of coefficients that should be taken into account using exact 
diagonalization is $7.5966\times 10^{10}$. Finally, for the $^6$Li-$^{173}$Yb mixture with $N_L=5$, $N_{Yb}=2$ and $C=(15;10;10)$ 
the corresponding coefficients within ML-MCTDHX are $7680$ while the inclusion of $6.6111\times 10^{12}$ coefficients is needed 
for a full configuration interaction treatment. 

Finally, let us also briefly showcase the numerical convergence of our results with respect to an increasing number 
of species functions $M$. 
We employ e.g. the time-evolution of the spectroscopic signal, $f(\Delta,t)_C$, at a certain rf-detuning $\Delta$ for the system consisting 
of $N_L=5$ and $N_K=2$ fermions. 
To infer about convergence we calculate the deviation of $f(\Delta,t)_{C'}$ between 
the $C'=\left(15;12;8\right)$ and other numerical configurations $C=\left(M;12;8\right)$, namely
\begin{equation}
\delta f(\Delta, t)_{C,C'} =f(\Delta,t)_{C'} -f(\Delta,t)_{C}. \label{converg_test} 
\end{equation} 

Figure \ref{Fig:convergence_figure} presents $\delta f(\Delta,t)_{C,C'}$ for the case of $N_L=5$ and $N_K=2$ at $g=5$ when considering a pulse 
characterized by a detuning $\Delta=2.2$ and $\Delta=2.5$ respectively. 
Recall that these values of $\Delta$ lie in the vicinity of the first two energetically lowest lying resonances of the rf spectrum discussed in 
the main text, see also Fig. \ref{Fig:1} (c).
Evidently, a systematic convergence of $\delta f(\Delta,t)_{C,C'}$ is achieved for both $\Delta=2.2$ and $\Delta=2.5$. 
For instance, comparing $\delta f(\Delta,t)_{C,C'}$ at $\Delta=2.2$ between the $C'=(15;12;8)$ and $C=(14;12;8)$ [$C=(10;12;8)$] approximations we can infer that 
the corresponding relative difference lies below $0.01\%$ [$0.1\%$] throughout the evolution, see Fig. \ref{Fig:convergence_figure} (a). 
Also, as illustrated in Fig. \ref{Fig:convergence_figure} (b) for $\Delta=2.5$ the corresponding $\delta f(\Delta,t)_{C,C'}$ between the configurations $C'=(15;12;8)$ 
and $C=(14;12;8)$ [$C=(10;12;8)$] shows a deviation which reaches a maximum value of the order of $0.15\%$ [$0.4\%$] at large pulse times. 
Finally, we note that a similar analysis has been performed for all other rf-detunings $\Delta$ and interspecies interaction strengths shown in the main text 
and found to be converged (results not shown here for brevity).

\vspace{0.1cm}
\begin{acknowledgements}
The authors acknowledge fruitful discussions with A. G. Volosniev, N. Zinner and A. Recati. 
G.M.K and P.S. acknowledge the support by the cluster of Excellence
'Advanced Imaging of Matter' of the Deutsche Forschungsgemeinschaft (DFG) - EXC 2056 - project ID 390715994.
\vspace{0.1cm}
G.C.K, S.I.M. and G.M.K. contributed equally to this work.

\end{acknowledgements}

{}

\begin{thebibliography}{60}
\bibitem{g1d} Note that the effective interaction strength is given in terms of the scattering length $a_0$
as $g_{1D}=\big(2\pi \hbar^2 a_0 \left(m_{L}+m_K\right)\big)/\left(m_{L} m_K\right)$.
\bibitem{comment3} For $g=1.5$ the corresponding separation of the spectral lines is suppressed 
resulting to a broadened central peak and a single and sharper side peak when compared to the $g=5$ case.
\end{thebibliography}
\end{document}